\documentclass[aps,pra,twocolumn,showpacs,amsfonts,amssymb]{revtex4}

\usepackage{xspace,psfig,epsfig,amsfonts,amssymb,color}

\begin{document}

\newcommand{\beq}{\begin{equation}} \newcommand{\eeq}{\end{equation}}
\newcommand{\bqa}{\begin{eqnarray}} \newcommand{\eqa}{\end{eqnarray}}
\newcommand{\nn}{\nonumber} \newcommand{\nl}[1]{\nn \\ && {#1}\,}
\newcommand{\erf}[1]{Eq.~(\ref{#1})}
\newcommand{\erfs}[2]{Eqs.~(\ref{#1})--(\ref{#2})}
\newcommand{\crf}[1]{Ref.~\cite{#1}} 
\newcommand{\dg}{^\dagger}
\newcommand{\rt}[1]{\sqrt{#1}\,}
\newcommand{\smallfrac}[2]{\mbox{$\frac{#1}{#2}$}}
\newcommand{\half}{\smallfrac{1}{2}}
\newcommand{\bra}[1]{\langle{#1}|} \newcommand{\ket}[1]{|{#1}\rangle}
\newcommand{\ip}[2]{\langle{#1}|{#2}\rangle}
\newcommand{\sch}{Schr\"odinger} \newcommand{\hei}{Heisenberg} 
\newcommand{\bl}{{\bigl(}} \newcommand{\br}{{\bigr)}}
\newcommand{\ito}{It\^o} \newcommand{\str}{Stratonovich}
\newcommand{\sq}[1]{\left[ {#1} \right]} 
\newcommand{\cu}[1]{\left\{ {#1} \right\}} 
\newcommand{\ro}[1]{\left( {#1} \right)}
\newcommand{\an}[1]{\left\langle{#1}\right\rangle}
\newcommand{\implies}{\Longrightarrow} 
\newcommand{\tr}[1]{{\rm Tr}\sq{ {#1} }} 
\newcommand{\del}{\nabla} \newcommand{\du}{\partial} 
\newcommand{\dbd}[1]{{\partial}/{\partial {#1}}}
\newcommand{\tp}{^{\top}} 
\newcommand{\tbt}[4]{\left( \begin{array}{cc} {#1}& {#2} \\ {#3}&{#4} \end{array}\right)}

\title{Optimal control of entanglement via quantum feedback}

\author{Stefano Mancini}
\email{stefano.mancini@unicam.it}

\affiliation{Dipartimento di Fisica, Universit\`{a} di Camerino, 
I-62032 Camerino, Italy\\
and INFN, Sezione di Perugia, I-06123 Perugia, Italy}

\author{Howard M.\ Wiseman} 
\email{H.Wiseman@griffith.edu.au}

\affiliation{Centre for
  Quantum Computer Technology, Centre for Quantum Dynamics, School of Science, 
   Griffith University, Brisbane 4111 Australia}

\date{\today}

\begin{abstract}
It has recently been shown that finding the optimal measurement on the environment for 
stationary Linear Quadratic Gaussian control problems is a semi-definite program. We apply this technique to the control of the EPR-correlations between two bosonic modes interacting via a parametric Hamiltonian at steady state. The optimal measurement turns out to be nonlocal homodyne measurement --- the outputs of the two modes must be combined before measurement. We also find the optimal local measurement and control technique. This gives the same degree of entanglement but a higher degree of purity than the  local technique previously considered [S. Mancini, Phys. Rev. A {\bf 73}, 010304(R) (2006)].
\end{abstract}

\pacs{03.67.Mn, 02.30.Yy, 42.50.Lc, 42.50.Dv}

\maketitle

\section{Introduction}
\label{sec:intro}

Quantum feedback control is a well-established theoretical technique for stabilizing an open quantum
 system in a state with certain desired properties \cite{Bel87,Wis94a,Doh99}. The basic idea is to use the information that leaks from the system into a bath to undo the undesirable effects of coupling to this, or other, baths. Notable examples include protecting a ``\sch\ cat'' superposition \cite{HorKil97}, correcting errors in encoded quantum information \cite{MabZol96,AhnWisMil03},
maintaining a two-level atom in an arbitrary state \cite{WisManWan02},  deterministically producing entangled states of spins \cite{ThoManWis02,StoHanMab04} (which has been experimentally demonstrated \cite{Ger04}), and cooling various systems  to (close to) their ground states \cite{MVT98,MVT00,Hop03,Ste04}.

Recently, one of us started to consider the application of quantum feedback to control entanglement. Preliminary studies have been carried out for two interacting qubits \cite{MW05} and two interacting bosonic modes \cite{M06}, damped to independent baths. 
Physically, two damped and interacting bosonic modes could be realized by optical cavity modes coupled by a $\chi^{(2)}$ nonlinearity. 
Such a nonlinearity induces only a finite amount of entanglement between the modes in steady-state. By contrast, it was shown in Ref.~\cite{M06} that performing homodyne detection on the two outputs, and using these currents to modulate the (linear) driving of the two modes could, under ideal conditions, increase this entanglement without limit.

The quantum control problem of Ref.~\cite{M06} has, like many which have been considered  \cite{Bel87,Wis94a,Doh99,ThoManWis02,Ste04}, an analogy in the class of classical LQG problems 
\cite{Jac93}. That is, systems with Linear dynamics and a Linear map from inputs to outputs, 
an aim that can be expressed in terms of minimizing a Quadratic cost function, and Gaussian noise in the dynamics and the outputs. 
The quantum LQG problem has recently been analysed in detail by one of us and Doherty \cite{WD05}. In particular, it was shown in Ref.~\cite{WD05} that in the quantum case, there is an extra level of optimization that naturally arises: choosing the optimal {\em unravelling} (way to extract information from the bath) given a fixed system--bath coupling.

In this paper we reconsider the problem of Ref.~\cite{M06} from the perspective of Ref.~\cite{WD05}. We formulate the problem as an LQG control problem, and find the optimal unravelling. This is different from the unravelling used in Ref.~\cite{M06} (it requires interfering the two output beams prior to homodyne detection) and leads to greater entanglement for any strength of the nonlinearity. This shows the usefulness of the general techniques of Ref.~\cite{WD05}.

The paper is organized as follows. In Sec.~\ref{sec:QLQG} we summarize the general theory of quantum LQG control problems from Ref.~\cite{WD05} as needed for the current problem. 
In Sec.~\ref{sec:ent} the problem of maximizing the steady-state entanglement between  two bosonic modes interacting via parametric Hamiltonian is addressed and the optimal unravelling found. 
As stated above, this involves interfering the output beams from the two modes prior to detection, which could be regarded as a nonlocal measurement. 
In Sec.~\ref{sec:loc} we consider the constraint of local measurements (that is,
independent measurements on the two outputs). We consider a variety of measurement and feedback schemes, including homodyne and heterodyne detection, and find that two schemes, that considered 
in Ref.~\cite{M06}, and another (more symmetric) scheme, are the best. However, when it comes to 
the purity  of the stationary entangled state, considered in Sec.\ref{sec:coh}, the more symmetric scheme is superior. 
 Finally, Sec.\ref{sec:conclu} is for conclusions.


\section{Quantum LQG control problems}
\label{sec:QLQG}

\subsection{Continuous Markovian Unravellings of Open Systems}
\label{sec:unrav}

We consider open quantum systems whose average (that is, unconditional) evolution
can be described by an autonomous differential equation for the state matrix $\rho$. 
The most general such equation is the Lindblad master equation \cite{Lin76}
\begin{equation}
\dot{\rho}=-i[\hat H,\rho ]+ {\cal D}[\hat{\bf c}]\rho  \equiv {\cal L}_0 \rho
\label{eq:lindblad}
\end{equation}
Here $\hat H = \hat H\dg$ is the system Hamiltonian 
(we use $\hbar=1$ throughout the paper), while $\hat{\bf c}$
is a vector of operators $\hat{\bf c} =(\hat{c}_1,\cdots,\hat{c}_{L})\tp $ 
that need not be Hermitian. The action of ${\cal D}[\hat{\bf c}]$
on an arbitrary operator $\rho$ is defined by 
\beq {\cal D}[\hat{\bf c}]\rho \equiv \sum_{l=1}^L \left[ \hat{c_l}\rho \hat{c}_l^{\dagger }-\frac{1}{2}\left( 
\hat{c}_l^{\dagger}\hat{c}_l\rho +\rho \hat{c}_l^{\dagger }\hat{c}_l\right)\right].  
\eeq
Master equations of this form can typically be derived if the system is coupled weakly
to an environment that is large (i.e. with dense energy levels). Under these conditions, 
 it is possible to measure the environment continually on a
time scale much shorter than any system time of interest. This {\em
  monitoring} yields information about the system, producing a
stochastic {\em conditional} system state $\rho_{\rm c}$ that {\em on
  average} reproduces the unconditional state $\rho$. That is, the
master equation is {\em unravelled} into stochastic quantum
trajectories \cite{Car93b}, with different measurements on the
environment leading to different unravellings.
 
For the purposes of this paper we can restrict to unravelings that 
 yield an evolution for $\rho_{\rm c}$ that is continuous and
 Markovian. In that case, it must be of the form \cite{WisDio01}
\begin{eqnarray}
  d{\rho}_{\rm c} &=& 
{\cal L}_0 \rho_{\rm c}dt + 
d {\bf z}\dg (t) \Delta_{\rm c} \hat{\bf c} \rho_{\rm c}  +\rho_{\rm c} \Delta_{\rm c}
\hat {\bf c}^{\dagger } d{\bf z}(t) .\label{eq:sme}
\end{eqnarray}
Note that here the $\dagger$ indicates transpose ($\top$) of the
vector and Hermitian adjoint of its components. We are also using the notation
$\Delta_{\rm c} \hat {o} \equiv \hat o - \an{ \hat o}_{\rm c}$, where
$\an{ \hat o}_{\rm c} \equiv {\rm Tr}[\rho_{\rm c} \hat{o}]$. Finally, 
 we have introduced a vector $d{\bf z} = (dz_1, \cdots ,dz_{L})\tp$ of
infinitesimal complex Wiener increments \cite{Gar85}. It satisfies
E$[d{\bf z}] = 0$, where E denotes expectation value, and for 
 efficient detection has the
correlations \cite{WisDio01}
 \beq \label{noisepower} d{\bf z} d{\bf z}\dg =   I
 dt , \;\; d{\bf z} d{\bf z}\tp =    \Upsilon dt.  \eeq 
 Here $\Upsilon$ is a symmetric complex matrix, which is constrained only by the condition 
 $U\geq 0$ where $U$ is the  {\em unravelling matrix}
 \begin{equation}
U \equiv  \frac{1}{2}\left( 
\begin{array}{cc}
I+\text{Re}\left[ \Upsilon \right]  & \text{Im}\left[ \Upsilon\right]  \\ 
\text{Im}\left[ \Upsilon\right]  & I-\text{Re}\left[ \Upsilon\right] 
\end{array}
\right).
\end{equation}

 Equation (\ref{eq:sme}) describes a quantum diffusion process for the conditional
 state $\rho_{\rm c}$. Equations of this form were first written down by Belavkin \cite{Bel87},
 and were derived independently by Carmichael \cite{Car93b} to describe homodyne detection
 in quantum optics. The measurement results upon which the evolution of $\rho_{\rm c}$ is
conditioned is a vector of complex functions 
\beq \label{defJz} {\bf
  J}\tp dt =\an{\hat{\bf c}\tp  + \hat{\bf c}\dg \Upsilon}_{\rm
  c}dt + d{\bf z}\tp .  \eeq Following the terminology from quantum optics,
 we will call  ${\bf J}$ a current.

 \subsection{Linear Systems} \label{sec:LinDyn}

 We now specialize to systems of $N$ degrees of freedoms, with the
 $n$th described by the canonically conjugate pair obeying the
 commutation relations $[\hat q_{n},\hat p_{n}] = i $. Defining a
 vector of operators \beq \hat \mathbf{x}=\left( \hat q_{1},\hat
   p_{1},...,\hat q_{N},\hat p_{N}\right) \tp , \eeq we can write
\begin{equation} 
\label{defsigma}
 \left[\hat x_{n},\hat x_{m}\right] =i  \Sigma _{nm},
\end{equation}
 where $\Sigma$ is the $(2N)\times(2N)$  symplectic matrix 
 \beq \Sigma=\bigoplus _{n=1}^{N}  \left(
\begin{array}{cc} 0 & 1 \\ 
-1 & 0 \end{array} \right)  = \Sigma^* = - \Sigma\tp = - \Sigma^{-1}.
\eeq 

  For a system with such a phase-space structure we can define 
a Gaussian state as one with a Gaussian Wigner function \cite{GarZol00}. 
We write the  mean vector as $\an{ \hat {\bf x}}$ and its covariance matrix as $V$:
\beq V_{nm}=\left( \langle \Delta \hat x_{n}\Delta \hat
    x_{m}\rangle +\langle \Delta \hat x_{m}\Delta \hat x_{n}\rangle
  \right) /2.\eeq
For these to define a quantum states, the necessary and sufficient condition 
is that \cite{Hol75} 
\begin{equation}
V+i  \Sigma /2\geq 0.
\label{eq:uncert} \label{GSLMI}
\end{equation}

To obtain linear dynamics for our system in phase-space we require 
that $\hat H$ be quadratic, and $\hat{\bf c}$ linear, in $\hat{\bf
  x}$:
\begin{equation}
\hat H=(1/2)\hat \mathbf{x}\tp G\hat \mathbf{x} -\hat \mathbf{x}\tp \Sigma B\mathbf{u}(t)  ,\;\;\hat {\bf c} = \tilde C \hat {\bf x},
\end{equation}
where $G$ is real and symmetric and $B$ is real. The second term in
$\hat H$ is linear in $\hat {\bf x}$ to ensure a linear map between
the time-dependent classical input ${\bf u}(t)$ to the system and the
output current ${\bf J}(t)$.  For such a system, the unconditional
master equation (\ref{eq:lindblad}) has a Gaussian state as its solution,
with the following moment equations
\begin{eqnarray}
{d\langle \mathbf{\hat{x}}\rangle}/dt  &=&  A\langle \hat \mathbf{x}\rangle +B\mathbf{u}(t) ,  \\
 d{V}/dt &=&AV+VA\tp +D. \label{dVdt}
\label{eq:dyn}
\end{eqnarray}
Here $A=\Sigma ( G + \text{Im}[ \tilde{C}\dg \tilde{C} ] )$ and $D=
  \Sigma \text{Re}[ \tilde{C}\dg \tilde{C}] \Sigma\tp$. 

 For  {\em conditional} evolution of linear quantum systems 
it is convenient to recast the complex current ${\bf J}$ of \erf{defJz} as a real current with
{\em uncorrelated} noises: 
as opposed to the complex current ${\bf J}$ with (in general) correlated noises:
\beq\label{eq:y} 
{\bf y} \equiv (  U)^{-1/2} \left(
\begin{array}{c}
\text{Re}\left[ {\bf J} \right]  \\ 
\text{Im}\left[ {\bf J} \right] 
\end{array}
\right) = C\langle \mathbf{\hat x}\rangle +\frac{d\mathbf{w}}{dt}.
\end{equation}
Here $ C = 2 (U )^{1/2} \bar{C}$, where 
\beq
\bar{C}\tp \equiv 
\left( \text{Re}[ \tilde{C}\tp] , \text{Im}[ \tilde{C}\tp] \right),
\eeq
 while $d\mathbf{w}$ is a vector of real Wiener increments satisfying $d\mathbf{w}d\mathbf{w}\tp =Idt$. For linear systems   
this conditional state $\rho_{\rm c}$  from  \erf{eq:sme} is Gaussian, 
 with the  conditional moment equations \cite{WD05}
\begin{eqnarray}
d\langle \hat \mathbf{x}\rangle_{\rm c}=\left[ A \langle \hat \mathbf{x}\rangle_{\rm c} +B\mathbf{u(
}t\mathbf{)}\right] dt
+ \left( V_{\rm c}C\tp +  \Gamma\tp \right) d\mathbf{w}  \label{eq:kfilt1}\\
\dot{V}_{\rm c} =AV_{\rm c}+V_{\rm c}A\tp +D 
-(V_{\rm c}C\tp + \Gamma\tp )(CV_{\rm c}+ \Gamma) . \label{eq:kfilt2} \label{MRE}
\end{eqnarray}
 Here $\Gamma = - (  U)^{1/2}S\bar{C}\Sigma\tp$, where 
\beq S= {\tbt{0}{I}{-I}{0}}.
\eeq

Note that the equation for $V_{\rm c}$ is deterministic. In many situations (including those
 considered later in this paper), \erf{MRE} has a unique steady-state solution \cite{WD05}
 which moreover is a so-called {\em stabilizing solution} \cite{Zhou96}. 
 We will notate such a solution $V_{\rm c}^{\rm ss}$ as $W_U$ to emphasize that it depends upon the unraveling $U$:
 \beq \label{AMRE}
0 =\Omega  W_U+W_U \Omega \tp -W_UC\tp CW_U  + E E\tp.
\eeq
Here $\Omega  = \Sigma[ G + \bar{C}\tp S(2U-I)\bar{C}]$ is Hamiltonian drift, while
$E=  \Sigma C\tp/2$ manifests the measurement back-action
noise resulting from having ${\bf y} \propto C\hat{\bf x}$. 
The set of $W_U$s, for all possible unravellings $U$ (including ineffecient monitoring), is the 
 set of real symmetric matrices satisfying the two linear matrix inequalities 
 $W_U+i  \Sigma /2\geq 0$ and 
\beq \label{LMI:cond}
D+A W_U + W_U A\tp  \geq 0.
\eeq
This is shown in Ref.~\cite{WD05}. Moreover,  given a $W_U$ that satisfies these inequalities, 
 a (not necessarily unique) unravelling $U$ that will generate it can be found by solving 
\beq \label{LME}
  R\tp U R = D+A W_U + W_U A\tp  ,
\eeq
where $R = 2\bar{C}W_U/  + S\bar{C}\Sigma$. 

\subsection{Optimal Quantum Control}
\label{sec:OQC}

In feedback control,  ${\bf u}(t)$
depends on the history of the measurement record ${\bf y}(s)$ for $s<t$.   
The typical aim of control over some interval $[t_0,t_1]$ is to minimize the expected value of a {\em cost function} \cite{Jac93}, the integral of the sum of positive functions of ${\bf x}(t)$ and ${\bf u}(t)$ for $t_0<t<t_1$. We are interested in the special case of Linear-Quadratic-Gaussian (LQG) control \cite{Jac93}: a {\em linear} system with a {\em quadratic} cost function and having {\em Gaussian} noise. For an LQG control problem it can be shown that the optimal ${\bf u}$ is linear in the phase-space mean: 
\beq
{\bf u}(t) = -K(t) \an{\hat{\bf x}}_{\rm c}(t),
\eeq
where the matrix $K(t)$ 
can be determined from $A$, $B$, and the cost functions. It 
is independent of $D$, $C$, and $\Gamma$.

In this paper we are concerned only with the properties of the system at steady-state, 
so our aim is to minimize $m = {\rm E}[h]$ in the limit $t_1\to\infty$, where
\beq \label{quadcost}
h = \an{ \hat{\bf x}\tp P \hat{\bf x} }_{\rm c},
\eeq
with  $P\geq 0$.  Note that in steady state 
\beq \label{usefulforMarkov}
{\rm E}_{\rm ss}[\an{ \hat{\bf x}\tp P \hat{\bf x} }_{\rm c} ] = {\rm tr}[W_U P] + 
{\rm E}_{\rm ss}[\an{\hat{\bf x}}_{\rm c}\tp P\an{\hat{\bf x}}_{\rm c}].
\eeq
Assuming (as is the case in our system) a stabilizing solution $W_U$ plus  
control over all relevant degrees of freedoms of the system (as will be the case 
if $B$ is invertible), 
the control can always be chosen to set $\an{\hat{\bf x}}_{\rm c}\to 0$, so that 
\beq \label{tracePV}
m_{\rm opt} = {\rm tr} [P W_U]. 
\eeq

For such systems it turns out \cite{WD05} that the same result (that is, $\an{\hat{\bf x}}_{\rm c}\to 0$) 
can always be achieved with Markovian feedback as introduced by Wiseman and Milburn \cite{WisMilFeedback}. This is a much simpler form of feedback; for a general linear system it means 
\beq
\mathbf{u}(t)  = F(t) \mathbf{y}(t)
\eeq
If $F$ is  time-independent, the 
 average evolution of the system is described simply by modifying the drift and diffusion matrices to 
 \bqa
A' &=& A+BFC, \label{Ap} \\
D' &=& D + BFF\tp B\tp + BF\Gamma + \Gamma\tp F\tp B\tp .  \label{Dp}
\eqa
 With $B$ invertible it can be shown that the optimal choice (which makes $\an{\hat{\bf x}}_{\rm c}\to 0$) is $BF = - W_UC\tp  - \Gamma\tp$. 


\section{Controlling Entanglement} 
\label{sec:ent}

We now specialize to the system examined in Ref.~\cite{M06}: 
a non-degenerate parametric oscillator \cite{reid} where two damped bosonic
modes $c_1$ and $c_2$ interact through a $\chi^{(2)}$ optical nonlinearity.
Treating the pump mode classically, this results in a quadratic Hamiltonian for 
the two modes.  The master equation is  
\begin{equation}
 \dot\rho=-i\left[\hat{H},\rho\right]+{\cal D}[\hat{c}_1]\rho+{\cal D}[\hat{c}_2]\rho\,,
\end{equation}
where 
\begin{equation}\label{Hpo}
\hat{H}=i\chi\left(\hat{c}_1^{\dag}\hat{c}_2^{\dag}-\hat{c}_1\hat{c}_2\right)=\chi\left(\hat{q}_1\hat{p}_2+\hat{q}_2\hat{p}_1\right)\,.
\end{equation}
Here $\chi$ is the coupling constant, proportional to the $\chi^{(2)}$ 
coefficient and the amplitude of the pump. We have also defined quadratures for the 
two modes via $\hat{c}_j = (\hat q_j + i \hat p_j)/\sqrt{2}$. 

This system fits within the general model described above, 
with $N=L=2$. Defining 
\begin{equation}
\hat{\bf x}^T\equiv(\hat{q}_1,\hat{p}_1,\hat{q}_2,\hat{p}_2)^T,
\end{equation}
we have 
\begin{equation}
G=\left(\begin{array}{cccc}
0&0&0&\chi\\
0&0&\chi&0\\
0&\chi&0&0\\
\chi&0&0&0
\end{array}\right)\,
\end{equation}
and
\begin{equation}
{\widetilde C}=\frac{1}{\sqrt{2}}\left(\begin{array}{cccc}
1&i&0&0\\
0&0&1&i
\end{array}\right).
\end{equation}

From the above theory we obtain
\beq
{\overline C}=\frac{1}{\sqrt{2}}\left(\begin{array}{cccc}
1&0&0&0\\
0&0&1&0\\
0&1&0&0\\
0&0&0&1
\end{array}\right)\,,
\end{equation}
\begin{equation}
A=\left(\begin{array}{cccc}
-\frac{1}{2}&0&\chi&0\\
0&-\frac{1}{2}&0&-\chi\\
\chi&0&-\frac{1}{2}&0\\
0&-\chi&0&-\frac{1}{2}
\end{array}\right)\,,
\eeq
\beq
D=\frac{1}{2}\left(\begin{array}{cccc}
1&0&0&0\\
0&1&0&0\\
0&0&1&0\\
0&0&0&1
\end{array}\right)\,.
\end{equation}

For $\chi < 1/2$, the system has a stationary state with mean zero and covariance matrix $V$ 
which can be found by setting $dV/dt$ to zero in \erf{dVdt}. 
The result can be written in terms of  
$2\times 2$ submatrices as 
\begin{equation}\label{Vst}
V=\left(
\begin{array}{cc}
\gamma&\sigma
\\
\sigma^T&\gamma
\end{array}\right)\,,
\end{equation}
where the matrix elements of $\gamma$ and $\sigma$ are 
\begin{eqnarray}
\gamma_{qq}&=&\gamma_{pp}=\frac{1}{2}\left(\frac{1}{1-4\chi^2}\right)\,,
\label{g11}\\
\gamma_{qp}&=&\gamma_{pq}=0\,,\\
\label{g12}
\sigma_{qq}&=&-\sigma_{pp}=\frac{\chi}{1-4\chi^2}\,,
\label{s11}\\
\sigma_{qp}&=&\sigma_{pq}=0\,,
\label{s12}
\end{eqnarray}

Since the steady state is Gaussian it is completely characterized by 
the correlation matrix (\ref{Vst}).
Then, its degree of entanglement can be quantified by means of the 
logarithmic negativity \cite{vidwer02}
\begin{equation}\label{L}
L\equiv\left\{
\begin{array}{ccr}
-\log\left(2\tilde{\zeta}_{-}\right)&\quad&{\rm if}\;\tilde\zeta<1
\\
0&\quad&{\rm otherwise}
\end{array}
\right.\,,
\end{equation}
where  
\begin{equation}\label{z}
\tilde\zeta_{-}\equiv\sqrt{\left(\det\gamma-\det\sigma\right)
-\sqrt{\left(\det\gamma-\det\sigma\right)^2-\det V}}
\end{equation}
 is the lowest symplectic eigenvalue of the partial transposed Gaussian state characterized by $V$. 
The quantity $L$ is represented by the curve $d)$ in Fig.\ref{figL}.
It is nonzero for all $\chi>0$, and is finite even as $\chi \to 1/2$. That is, the damping channels 
degrade the system state, preventing it from becoming maximally entangled 
like that of Ref.~\cite{epr}. 

A Gaussian with a covariance matrix of the form of \erf{Vst} is entangled if and only if 
 the variance of the mixed quadratures  
\begin{equation}
\hat{x}_j(\theta)=\cos\theta \hat{q}_j+\sin\theta \hat{p}_j
\end{equation}
is less than the the vacuum 
fluctuation level of unity \cite{sim,duan}. It is easy to calculate that 
\begin{equation}\label{var1-2}
\langle \left(\hat{x}_1(\theta)+\hat{x}_2(\pi-\theta)\right)^2\rangle
=\frac{1}{1+2\chi}\,.
\end{equation}
We see that the variances (\ref{var1-2}) go below $1$ as soon as $\chi>0$, from which we infer 
 entanglement. Since we must have  
$\chi<1/2$, the variances
(\ref{var1-2}) are limited from below by $1/2$. This is even though the 
variance in $\hat x_j(\theta)$, for either $j$ and for all $\theta$, are  
unbounded as $\chi \to 1/2$. This again shows that the stationary state
has only a finite amount of entanglement, and is not pure. For states of this form,
the log-negativity is in fact a simple function of the above variance:
\beq
L = -\log_2[1/(1+2\chi)].
\eeq


\subsection{Optimal Measurement and Control}

As we have seen, entanglement is manifest in the squeezing  of the quadratures in \erf{var1-2}, for all $\theta$. 
Thus, as an aim for the feedback, we can choose the minimization of 
\beq
\int \frac{d\theta}{2\pi} \langle \left(\hat{x}_1(\theta)+\hat{x}_2(\pi-\theta)\right)^2\rangle .
\eeq
in steady state. This evaluates to
\beq \label{qmpp}
\langle(\hat{q}_1 - \hat{q}_2)^2\rangle/2 + \langle(\hat{p}_1 + \hat{p}_2)^2\rangle/2
\eeq
This is exactly of the form of \erf{quadcost}, with 
\begin{equation}
P=\frac{1}{2}\left(\begin{array}{cccc}
1&0&-1&0\\
0&1&0&1\\
-1&0&1&0\\
0&1&0&1
\end{array}\right).
\end{equation}
From the symmetry of the problem, we  can assume that the optimal conditional covariance matrix shares 
the same structure as the unconditional matrix, namely:
\begin{equation}
W_U=\left(\begin{array}{cccc}
\alpha&0&\beta&0\\
0&\alpha&0&-\beta\\
\beta&0&\alpha&0\\
0&-\beta&0&\alpha
\end{array}\right)\,.
\end{equation}
Thus the quantity to be minimized is
\begin{equation}
m={\rm Tr}\left[PW_U\right]=2\left(\alpha-\beta\right)\,.
\end{equation}

\begin{figure}
\begin{center}
\includegraphics[width=0.45\textwidth]{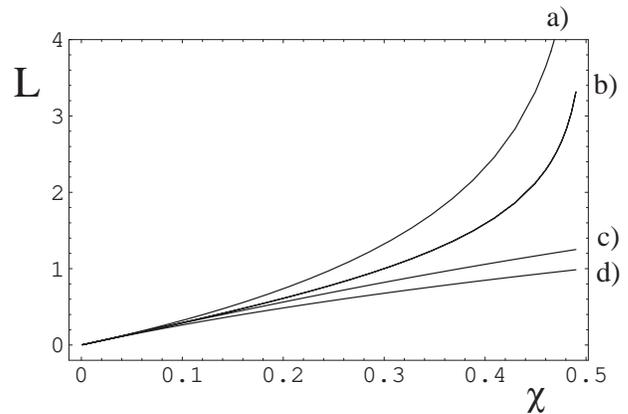}
\end{center}
\vspace{-0.5cm}
\caption{\label{figL} The logarithmic negativity $L$ of the steady-state quantum state of the 
non-degenerate OPO, versus the optical nonlinearity strength $\chi$, for: 
a) optimal (nonlocal) measurement and feedback of Sec.~III~A; b) optimal local  
measurement (homodyne) and feedback of Sec.~IV~A; c) non-optimal local measurement 
(heterodyne) and feedback of Sec.~IV~B; and d) no feedback.
}
\end{figure}

We have to find the minimum of $m$ constrained by 
$W_U+(i/2)\Sigma\ge 0$ and $D+AW_U+W_UA^T\ge 0$.
In terms of $\alpha$ and $\beta$ this becomes
\begin{eqnarray}
&&\min\left(\alpha-\beta\right)\,,\\
&&\alpha-\frac{1}{2}\sqrt{1+4\beta^2}\ge 0\,,\label{cond1}\\
&&\frac{1}{2}-\left(\alpha\pm\beta\right)\left(1\mp2\chi\right)\ge 0\,.
\label{cond2}
\end{eqnarray}
Taking $\alpha=\frac{1}{2}\sqrt{1+4\beta^2}$ we get $m=2\left(\sqrt{1+4\beta^2}-2\beta\right)$, 
which decreases monotonically with $\beta$. But from the condition (\ref{cond2}) we obtain 
\begin{equation}
\frac{1}{2}-\frac{1}{2}\left(\sqrt{1+4\beta^4}\pm2\beta\right)\left(1\mp2\chi\right)\ge 0\,.
\end{equation}
That is, 
\begin{equation}
\beta\le\chi\frac{1-\chi}{1-2\chi}\,,\qquad\left(0\le\chi<\frac{1}{2}\right)\,.
\end{equation}
Thus, choosing $\beta=\chi\frac{1-\chi}{1-2\chi}$ and $\alpha=\frac{1}{2}\sqrt{1+4\beta^2}$,
we obtain the minimum $m_{\rm opt}$. In this case the logarithmic negativity takes a simple analytical form
\beq
L_{\rm opt}= -\log_2[m_{\rm opt}] = -\log_2[1-2\chi]\, .
\label{Lfbopt}
\eeq
This is represented by the curve $a)$ of Fig.\ref{figL}. Note that this is unbounded as $\chi \to 1/2$.  

\begin{figure}
\begin{center}
\includegraphics[width=0.45\textwidth]{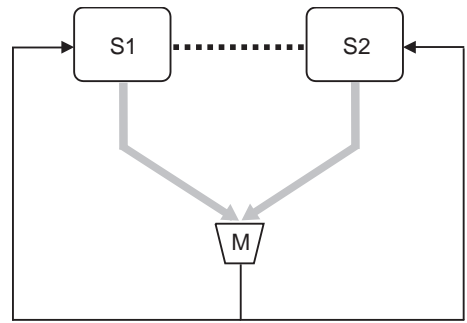}
\end{center}
\vspace{-0.5cm}
\caption{\label{diag_a} 
Schematic representation of feedback action based on nonlocal measurements.
$S1$ and $S2$ are the two interacting systems and $M$ is a common measurement box.
}
\end{figure}

If now we wish to know how to achieve this optimal result, we can use Eq.~(\ref{LME}) to get
\begin{equation}
U=\frac{1}{2}\left(\begin{array}{cccc}
1&-1&0&0\\
-1&1&0&0\\
0&0&1&1\\
0&0&1&1
\end{array}\right)\,.
\end{equation}
Since $U=U^{1/2}$ we can easily derive the matrix $C$ of Eq.~(\ref{eq:y})
\begin{equation}
C=\frac{1}{\sqrt{2}}
\left(\begin{array}{cccc}
1&0&-1&0\\
-1&0&1&0\\
0&1&0&1\\
0&1&0&1
\end{array}\right)\,.
\end{equation}
This tells us that the optimal unraveling is the measure of $\hat{q}_1-\hat{q}_{2}$ (first and second rows)
and $\hat{p}_1+\hat{p}_{2}$ (third and fourth rows). Intuitively this makes sense, as it is the variances of these
quantities that we wish to minimize, from \erf{qmpp}. Note however that these are 
 \emph{nonlocal} measurements in the sense that they involve combinations of observables belonging to the 
 different subsystems. That is, the output beam from mode 1 must be mixed at a beam splitter with the output beam from 
 mode 2, and then the two beam-splitter outputs subject to homodyne detection. One of these detections
can measure the output quadrature corresponding to $\hat q_1 - \hat q_2$, and the other can measure 
that corresponding to $\hat p_1 + \hat p_2$. 
This situation of nonlocal measurements is schematically represented in Fig.\ref{diag_a}.

\section{Local Feedback Action}
\label{sec:loc}

Having shown that the optimal control protocol involves a nonlocal measurement, 
it is natural to ask how much improvement this offers over protocols involving only local
measurements on the two subsystems. The latter would be easier to implement experimentally, especially if 
the two modes had significantly different frequencies. In previous work \cite{M06},
only local measurements were considered.  


\subsection{Single quadrature measurements}

We begin by considering a homodyne measurement of the output beam of each mode. 
From Eq.~(\ref{var1-2}) we see that there are no preferred quadratures to 
be measured provided that their angles sum up to $\pi$.  
Without loss of generality we can assume to measure $\hat{q}_1$ and $\hat{q}_2$. In terms of the parameters 
of Sec.~\ref{sec:unrav} we require 
\begin{equation}
\Upsilon={\rm diag}(1,1),
\end{equation}
so that $J_1\propto q_1$ and $J_2\propto q_2$.

The Hamiltonian term $ -\hat \mathbf{x}\tp \Sigma BF\mathbf{y}$ would represents the feedback Hamiltonian. Since we measure $q_1$ and $q_2$, it is natural to act on the quadrature to $\hat q_1 - \hat q_2$, in order to minimize its variance.
That is, we choose the feedback to be proportional to the conjugate quadrature:
\begin{eqnarray}\label{HfbLF}
\hat{H}_{\rm fb}&=&\lambda_{-} \left[J_1(t)-J_2(t)\right]\times\left[\hat{p}_1-\hat{p}_2\right]\nonumber\\
&+&\lambda_{+} \left[J_1(t)+J_2(t)\right]\times\left[\hat{p}_1+\hat{p}_2\right].
\end{eqnarray}
Here $\lambda_{\pm}$ represents possible feedback strengths. 
Equation (\ref{HfbLF}) represents the most general feedback action that 
accounts for the symmetry between the two subsystems. Note that this feedback can be performed 
locally because the Hamiltonian (\ref{HfbLF}) contains no products of operators for both subsystems. 
However, in general it requires classical communication, so that the controller for mode 1 can apply a Hamiltonian 
proportional to $J_2$, and vice versa. \erf{HfbLF} is obtained by choosing the feedback driving like 
\begin{equation}
B F=\frac{1}{\sqrt{2}}\left(\begin{array}{cccc}
\lambda_{+}+\lambda_{-}&\lambda_{+}-\lambda_{-}&0&0\\
0&0&0&0\\
\lambda_{+}-\lambda_{-}&\lambda_{+}+\lambda_{-}&0&0\\
0&0&0&0\\
\end{array}
\right).
\end{equation}

\begin{widetext}

As consequence of feedback action the matrices $A$ and $D$ are modified according to \erf{Ap} and \erf{Dp} to
\begin{equation}\label{ALF}
A'=\left(\begin{array}{cccc}
-\frac{1}{2}+\lambda_{-}+\lambda_{+}&0&\chi-\lambda_{-}+\lambda_{+}&0
\\
0&-\frac{1}{2}&0&-\chi
\\
\chi-\lambda_{-}+\lambda_{+}&0&-\frac{1}{2}+\lambda_{-}+\lambda_{+}&0
\\
0&-\chi&0&-\frac{1}{2}
\end{array}\right),
\end{equation}
\begin{eqnarray}\label{N}
D'=\frac{1}{2}\left(\begin{array}{cccc}
(1-\lambda_{-}-\lambda_{+})^2+(\lambda_{-}-\lambda_{+})^2&0&2(1-\lambda_{-}-\lambda_{+})(\lambda_{-}-\lambda_{+})&0
\\
0&1&0&0
\\
2(1-\lambda_{-}-\lambda_{+})(\lambda_{-}-\lambda_{+})&0&(1-\lambda_{-}-\lambda_{+})^2+(\lambda_{-}-\lambda_{+})^2&0
\\
0&0&0&1
\end{array}\right)\,.
\end{eqnarray}
The stationary covariance matrix that results from these is of the form of \erf{Vst} with 
\begin{eqnarray}
\gamma_{qq}&=&\frac{-1+4(1+\chi)\lambda_{+}-2(1+2\chi)\lambda_{+}^2+\lambda_{-}^2(-2+4\chi+8\lambda_{+})-4\lambda_{-}(-1+\chi+4\lambda_{+}-2\lambda_{+}^2)}{2(1+2\chi-4\lambda_{-})(-1+2\chi+4\lambda_{+})}\,,
\label{g11mod}\\
\gamma_{qp}&=&\gamma_{pq}=0\,,
\label{g12mod}\\
\gamma_{pp}&=&\frac{1}{2}\left(\frac{1}{1-4\chi^2}\right)\,, \\
\label{g22mod}
\sigma_{qq}&=&\frac{\lambda_{-}^2(1-4\lambda_{+})-\lambda_{+}^2+4\lambda_{-}\lambda_{+}^2+\chi(-1+2\lambda_{-}-2\lambda_{-}^2+2\lambda_{+}-2\lambda_{+}^2)}{(1+2\chi-4\lambda_{-})(-1+2\chi+4\lambda_{+})}\,,
\label{s11mod}\\
\sigma_{qp}&=&\sigma_{pq}=0\,,
\label{s12mod}\\
\sigma_{pp}&=&-\frac{\chi}{1-4\chi^2}\,.
\label{s22mod}
\end{eqnarray}

\end{widetext}

We have maximized the logarithmic negativity (\ref{L}) over $\lambda_+$ and $\lambda_-$. 
with the constraints that $V$ be a stable solution to \erf{dVdt}. 
In the range $0<\chi< 1/2$, these constraints are
\beq
\lambda_{\pm}<\frac{1}{4}\mp\frac{\chi}{2}.
\eeq
We summarize the results by distinguishing four limit cases for which 
$L$ becomes dependent on a single parameter.

\begin{figure}
\begin{center}
\includegraphics[width=0.45\textwidth]{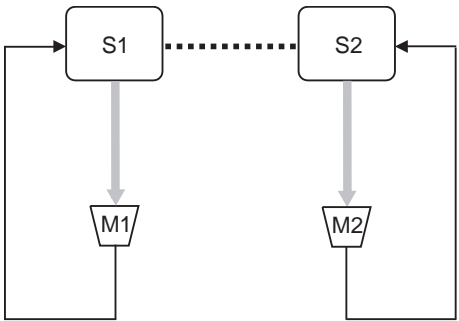}
\end{center}
\vspace{-0.5cm}
\caption{\label{diag_c} 
Schematic representation of feedback action based on local measurements and no classical communication.
$S1$ and $S2$ are the two interacting systems and $M1$, $M2$ are local measurement boxes.
}
\end{figure}

\begin{figure}
\begin{center}
\includegraphics[width=0.45\textwidth]{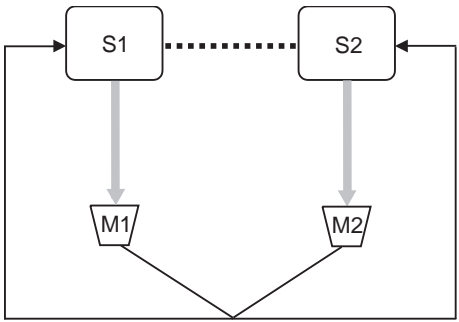}
\end{center}
\vspace{-0.5cm}
\caption{\label{diag_b} 
Schematic representation of feedback action based on local measurements supplied by classical communication.
$S1$ and $S2$ are the two interacting systems and  $M1$, $M2$ are local measurement boxes.
}
\end{figure}

\begin{description}
\item[i)]
If we set 
 $\lambda_{\pm}=\lambda$ we have a purely
local feedback, without classical communication. This is because the first (respectively second) current is used to control the first (respectively second) subsystem (see Fig.\ref{diag_c}).
This case does not show any improvement with respect to the no-feedback case; that is the optimal value of  parameter is $\lambda=0$ [it corresponds to curve $d)$ of Fig.\ref{figL}]. This is  because, with local measurements and 
no communication, the correlations between the two subsystems cannot be increased.

\item[ii)]
If we set  $\lambda_{-}=0$ and $\lambda_{+}=\lambda$ we do require classical communication (see Fig.\ref{diag_b}). 
However, also this case does not show improvement with respect to the no-feedback case; the optimal value of  parameter is $\lambda=0$ [it corresponds to curve $d)$ of Fig.\ref{figL}]. This is 
 because the corresponding feedback Hamiltonian is not effective acting on the antisqueezed quadrature $q_1+q_2$.

\item[iii)]
If we set $\lambda_{-}=\lambda$ and $\lambda_{+}=0$ we again require classical communication (see Fig.\ref{diag_b}). In this 
case the feedback Hamiltonian coincides with that used in Ref.~\cite{M06}.
 The optimal value of the feedback parameter is $\lambda=\chi$, 
and gives rise to a great improvement in the logarithmic negativity with respect to the no-feedback case
[it corresponds to curve $b)$ of Fig.\ref{figL}].
By approaching the instability point
$\chi\to 1/2$ the logarithmic negativity
increases indefinitely.

\item[iv)]
If we set  $\lambda_{\pm}=\pm\lambda$ we once again require classical communication 
 (see Fig.\ref{diag_b}).
The optimal value of parameter is $\lambda=-\chi$, 
and gives rise exactly to the same values of the logarithmic negativity as for the case {\bf iii)}
[thus corresponding to curve $b)$ of Fig.\ref{figL} too].
However, in Sec.~\ref{sec:coh} it will become clear that the case {\bf iv)} is superior to the case {\bf iii)} in other ways.

\end{description}

That case {\bf iv)} gives the best result is not surprising, since it gives rise to a feedback Hamiltonian that resemble that in Eq.~(\ref{Hpo}), once it is remembered that $J_1\propto q_1$ and $J_2\propto q_2$. Note that although in the cases {\bf iii)} and {\bf iv)}  the entanglement increases without bound as $\chi \to 1/2$, the log-negativity is still below that of 
the optimal nonlocal feedback for all values of $\chi$ as shown by Fig.\ref{figL}.


\subsection{Joint quadratures measurements}

Since \erf{qmpp} contains both $q$ and $p$, one might think that performing joint quadratures measurements in both subsystems would be an effective route to controlling entanglement. Of course it is not possible to measure both quadratures with perfect efficiency, but it is possible to measure each quadrature with an efficiency of $1/2$. This 
can be achieved by heterodyne measurement, for example \cite{WisMil93c}.  In terms of the parameters 
of Sec.~\ref{sec:unrav} we require $\Upsilon=0$ so that $J_1\propto c_1$ and $J_2\propto c_2$.

Bearing in mind the results of preceding subsection (that is, that scheme {\bf iv)} performed best) 
we restrict our consideration to 
feedback that gives rise to Hamiltonian resembling the one in \erf{Hpo}.
Hence we choose the feedback driving as
\begin{equation}
B F=\left(\begin{array}{cccc}
0&\mu&0&0\\
0&0&0&-\mu\\
\mu&0&0&0\\
0&0&-\mu&0
\end{array}
\right),
\end{equation}
corresponding to the feedback Hamiltonian 
\begin{equation}\label{HfbLFhet}
\hat{H}_{\rm fb}=-i\frac{\mu}{\eta} \left[\left(J_1(t)\hat{c}_2- J_1^{*}(t)\hat{c}_2^{\dag}\right)
+\left(J_2(t)\hat{c}_1- J_2^{*}(t) \hat{c}_1^{\dag}\right)\right].
\end{equation}
Here $\mu$ represents the feedback strength and $\eta\equiv 1/2$ accounts for the half unit efficiency.

\begin{widetext}

As consequence of feedback action, the matrices $A$ and $D$ are modified according to \erf{Ap}, \erf{Dp} to
\begin{equation}\label{ALFhet}
A'=\left(\begin{array}{cccc}
-\frac{1}{2}&0&\chi+\mu&0
\\
0&-\frac{1}{2}&0&-\chi-\mu
\\
\chi+\mu&0&-\frac{1}{2}&0
\\
0&-\chi-\mu&0&-\frac{1}{2}
\end{array}\right),
\end{equation}
\begin{eqnarray}\label{Nhet}
D'=\frac{1}{2}\left(\begin{array}{cccc}
1+2\mu^2&0&-2\mu&0
\\
0&1+2\mu^2&0&2\mu
\\
-2\mu&0&1+2\mu^2&0
\\
0&2\mu&0&1+2\mu^2
\end{array}\right)\,.
\end{eqnarray}
Proceeding as above,  the stationary covariance matrix elements resulting are given by
\begin{eqnarray}
\gamma_{qq}&=&\frac{-1+4\chi\mu+2\mu^2}
{2(-1+4(\chi+\mu)^2)}=\gamma_{pp}\,,
\label{g11modhet}\\
\gamma_{qp}&=&\gamma_{pq}=0\,,
\label{g12modhet}
\end{eqnarray}
\begin{eqnarray}
\sigma_{qq}&=&-\frac{\chi+2\chi\mu^2+2\mu^3}{-1+4(\chi+\mu)^2}=-\sigma_{pp}\,,
\label{s11modhet}\\
\sigma_{qp}&=&\sigma_{pq}=0\,,
\label{s12modhet}
\end{eqnarray}

\end{widetext}

We have maximized the logarithmic negativity (\ref{L}) over $\mu$
with the constraints that $V$ be a stable solution to \erf{dVdt}. 
In the range $0<\chi< 1/2$ these are
\beq
-\frac{1}{2}-\chi<\mu<\frac{1}{2}-\chi.
\eeq
We summarize the results hereafter.

\begin{description}

\item[v)]
The optimal value of the parameter $\mu$ is found to be $\mu=(-1-2\chi+\sqrt{1+4\chi^2})/2$. This gives rise to a small improvement in the logarithmic negativity with respect to the no-feedback case --- it corresponds to curve $c)$ of Fig.\ref{figL}.

\end{description}

Although this case does improve entanglement, 
it is not as good as the best homodyne scheme ${\bf iv)}$. 
This can be understood as follows.
In controlling a quantum system, one has always to reaach a 
tradeoff between information gain and introduced disturbance.
Heterodyne detection allows us to gain information about both system quadratures, in contrast to homodyne detection, at expenses of introducing more noise via the feedback.
In our system, it is apparent that a high degree of entanglement can be produced by 
controlling only one pair of quadratures, so the noise introduced by heterodyne-based feedback 
produces inferior performance relative to homodyne-based feedback. 
In other contexts (with other Hamiltonians) heterodyne-based feedback may outperform 
homodyne-based.


\section{Purity}
\label{sec:coh}

The fact that for optimal nonlocal feedback, and local feedback of cases {\bf iii)}, {\bf iv)}, the entanglement can increased without bound, means that feedback is able to recycle the information lost by the system into environment through the amplitude damping. However the EPR correlations  \cite{epr} imply not only an arbitrarily entangled, but also a pure state.  We now check 
what the purity of our stationary state is under the various feedback control schemes.

For a Gaussian state the von Neumann entropy can be written as \cite{holevo}
\begin{equation}\label{vonN}
S=g(\zeta_{+})+g(\zeta_{-})\,,
\end{equation}
where 
\begin{equation}
g(x)\equiv \left(x+\frac{1}{2}\right)\log \left(x+\frac{1}{2}\right)
- \left(x-\frac{1}{2}\right)\log \left(x-\frac{1}{2}\right)
\end{equation}
and
\begin{equation}\label{zpm}
\zeta_{\pm}\equiv\sqrt{\left(\det\gamma+\det\sigma\right)
\pm\sqrt{\left(\det\gamma+\det\sigma\right)^2-\det V}}
\end{equation}
are the symplectic eigenvalues of the Gaussian state characterized by $V$.

\begin{figure}
\begin{center}
\includegraphics[width=0.4\textwidth]{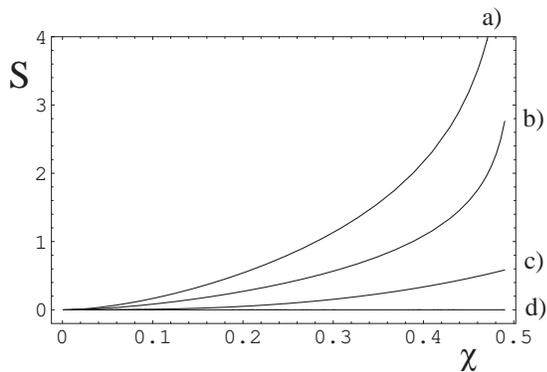}
\end{center}
\vspace{-0.5cm}
\caption{\label{figS} 
The von Neumann entropy $S$ of the steady-state quantum state of the 
non-degenerate OPO, versus the optical nonlinearity strength $\chi$, for: 
a) no feedback;
b) non-optimal local  
measurement (homodyne) and feedback of Sec.~IV~A;
c) non-optimal local measurement 
(heterodyne) and feedback of Sec.~IV~B;
d) optimal (nonlocal) measurement and feedback of Sec.~III~A and optimal local  
measurement (homodyne) and feedback of Sec.~IV~A.
}
\end{figure}

We have numerically evaluated the quantity (\ref{vonN}) for nonlocal and local feedback action and the results are shown in Fig.\ref{figS}.
The curve $a)$ corresponds to the worst cases {\bf i)}, and {\bf ii)} also corresponding to the no-feedback action. Below is the curve $b)$ corresponding to the case {\bf iii)} and showing that the state remains not pure.  In this case we have the mixedness, as well as the amount of entanglement, increasing by increasing $\chi$. The fact that they both increase indefinitely may sound strange. However, the limit $\chi\to1/2$ has to be taken with care, and the above results are justified by the fact that it allows infinite energy to come into the state.
The curve $c)$ corresponds to the case  {\bf v)}.

Finally, the curve $d)$ corresponds to optimal nonlocal feedback and case {\bf iv)}, thus showing that in such cases the entropy is always zero and the purity of the state is restored by the feedback action. 
These results (Fig.\ref{figS}) together with those of logarithmic negativity (Fig.\ref{figL}) clearly show the optimality of the feedback scheme {\bf iv)} among the local schemes, and the global optimality 
of the nonlocal scheme.


\section{Conclusions}
\label{sec:conclu}

Summarizing, we have found the optimal nonlocal feedback action as well as the optimal local one to control steady state EPR-correlations for two bosonic modes interacting via parametric Hamiltonian $\propto \chi$.
Both these actions allow one to produce arbitrary amounts of entanglement as $\chi \to 1/2$, although more in the former case. Moreover, they both do this while producing a pure state --- 
that is they permit us recover the coherence of our open quantum system. 
(Incidentally the possibility of coherence recovery by means of feedback was forecast in Ref.~\cite{mauro} for finite dimensional systems by an information theoretic approach.)

Our local feedback action requires only classical communication and Gaussian operations 
(linear displacements). This may appear to contradict the
impossibility to enhance (distill) entanglement by means of 
Gaussian LOCC stated in  Refs.\cite{eis}. 
The key point is that, in contrast with Refs.\cite{eis},
here the LOCC operations continuously happen while the entangling interaction is ``on".
Thus, the presented approach may shed some light on the subject
of entanglement distillation.

While we have used a semidefinite program to find the optimal measurement and 
feedback action for general LQG systems, 
to find the optimum local scheme we used simple optimization informed by 
the symmetries of the system. 
The question of defining an efficient  program to find the 
 optimal {\em local} measurement for feedback control of general LQG systems remains open.


\acknowledgments
SM acknowledge financial support from the 
European Union through the Integrated Project FET/QIPC ``SCALA". 
HMW is supported by the Australian Research Council and the Queensland Government.


\end{document}